\documentclass[journal]{IEEEtran}
\ifCLASSINFOpdf
   \usepackage[pdftex]{graphicx}
   \graphicspath{{../pdf/}{../jpeg/}}
   \DeclareGraphicsExtensions{.png,.jpeg,.png}
\else
   or other class option (dvipsone, dvipdf, if not using dvips). graphicx
   will default to the driver specified in the system graphics.cfg if no
   driver is specified.
   \usepackage[dvips]{graphicx}
   declare the path(s) where your graphic files are
   \graphicspath{{../eps/}}
   and their extensions so you won't have to specify these with
   every instance of \includegraphics
   \DeclareGraphicsExtensions{.eps}
\fi

\usepackage{amsmath}
\usepackage{booktabs}
\usepackage{multirow}
\begin{document}

%
\title{How to Eliminate Detour Behaviors in E-hailing? Real-time Detecting and Time-dependent Pricing}
%
%
%

\author{Qiong Tian, Yue Yang, Jiaqi Wen, Fan Ding and Jing He
\thanks{The authors are with the School of Economics and Management, Beihang University, 37 Xueyuan Road, Beijing, 100191, China PR (e-mail: tianqiong@buaa.edu.cn).
}}



%
%


\markboth{IEEE TRANSACTIONS ON INTELLIGENT TRANSPORTATION SYSTEMS}%
{Shell \MakeLowercase{\textit{et al.}}: Bare Demo of IEEEtran.cls for IEEE Journals}

%



\maketitle

\begin{abstract}
With the rapid development of information and communication technology (ICT), taxi business becomes a typical electronic commerce mode. However, one traditional problem still exists in taxi service, that greedy taxi drivers may deliberately take unnecessary detours to overcharge passengers. The detection of these fraudulent behaviors is essential to ensure high-quality taxi service. In this paper, we propose a novel framework for detecting and analyzing the detour behaviors both in off-line database and among on-line trips. Applying our framework to real-world taxi data-set, a remarkable performance (AUC surpasses 0.98) has been achieved in off-line classification. Meanwhile, we further extend the off-line methods to on-line detection, a warning mechanism is introduced to remind drivers and an excellent precision (AUC surpasses 0.90) also has arrived in this phases. After conducting extensive experiments to verify the relationships between pricing regulations and detour behaviors, some quantitative pricing suggestions, including rising base fare and reducing distance-based fare rate, are provided to eliminate detour behaviors from the long term.
\end{abstract}


%
\IEEEpeerreviewmaketitle

\section{Introduction}

\IEEEPARstart{W}{ithin} recent years, the accelerating development in sensing, communication, storage and computing technologies have changed transportation service in various aspects. One representative is the emerging of E-hailing platforms, including Uber, Lyft and Didi Chuxing, which have redefined the industry of taxi service. According to 2018's statistics in DiDi chuxing, there are over 100,000 taxis operating every day in Beijing, and about 70,000 in Shanghai. At present, the majority of operated taxis on E-hailing platforms are equipped with GPS navigation equipment, millions of taxis periodically upload their positions, directions, and speed as pervasive sensors of the road network each day, thereby creating a massive amount of trajectory data over time \cite{yin2014mining}. Therefore, gathering and analyzing these large-scale taxis’ GPS trajectories not only enable us to discovery the hidden “facts” about community dynamics and human behaviors \cite{yuan2014human,liu2013fraud}, but also show a great potential to revolutionize the services in various areas ranging from public safety, urban planning to transportation management\cite{zhang2011emergence,zhang2017correction}. In addition, lots of value-added applications, such as map service \cite{dai2015personalized}, point of interest recommending \cite{cheng2013you}, and spatial-temporal demand prediction \cite{yao2018deep}, could also benefit from big data analysis over taxis’ GPS data sets. 

For a long time, however, taxi service has been facing a typical challenge that greedy taxi drivers may deliberately take unnecessary detours to overcharge passengers, especially when passengers travel in unfamiliar cities. Currently, according to the feed-backs from passengers, large numbers of experienced staffs in customer service department are responsible for detecting taxi driving frauds via manually checking the corresponding taxis’ trajectories. Nonetheless, it is extremely difficult for these staffs to identify the suspicious taxis efficiently and precisely due to the exponentially increasing number of trips. Specifically, there also exists lots of frauds are not even noticed and perceived by passengers. In order to guarantee the quality of taxi service, it is crucial for these platforms to efficiently and automatically detect and disclose the detour trips from collected trajectories. Having discovered these mis-behaviors, the platforms further can take some measures, including giving a warning, fining a penalty and so on, to control and eliminate the detours' frequency in the future.

Most previous studies have proposed anomaly detection using trajectory data, which can be analyzed in several ways, including distance-based method \cite{knorr2000distance}, density-based method \cite{breunig2000lof}, similarity-based \cite{zhang2011ibat,chen2013iboat,zhou2016method} method and statistic-learning method \cite{li2007roam, sillito2008semi, xiao2015travel,dabiri2018inferring}. When applying these detection techniques to real-time framework, for incomplete GPS trajectory only can be collected, we should integrate a variety of other information, including dynamic route recommendations, estimated time arrival (ETA) and live traffic, to achieve accurate detection. On the other hand, since both demand and supply on E-hailing platforms have strong temporal patterns, in order to react to instantaneous imbalances between real-time demand and supply, dynamic pricing has been commonly used by these platforms. \cite{jiao2018investigating,guo2018modelling}. By contrast to fixed price or fixed wage, dynamic pricing mechanism has been proved to increase the platform’s profits \cite{cachon2017role} and avoid system inefficiency \cite{castillo2017surge}. Nevertheless, this pricing mechanism may also lead to some undesirable consequences such as detour frauds \cite{liu2017pricing} to some extent. 

In view of these challenges, in this paper we propose a novel detection framework to monitor the detour frauds both in off-line database and among on-line trips. Parallel to detection mechanism, a dynamic pricing suggestions have also been provided in the following phases to eliminate the detour behaviors on E-hailing platforms. In summary, the major contributions of our paper are showed below:

\begin{itemize}

  \item [(1)]
By extracting the distanced-based feature and time-based feature in off-line phases, the proposed method naturally falls into the category of supervised learning. Implementing the logistic regression model, our method has been proved achieving a remarkable performance by using millions of historical taxi data.

  \item [(2)]
Based on aforementioned off-line model, we further employ the estimated parameter to conduct an on-line detection and propose a real-time warning mechanism. Applying these technique to the on-line experiments, our proposed method ultimately arrives at an excellent precision for giving a warning to the anomalous drivers in real-time.

  \item [(3)]
From the perspective of the long term, we also conduct comprehensive discussions upon pricing regulation in E-hailing platforms. Consequently, several pricing suggestions have been given to motivate drivers to avoid detours. 

\end{itemize}

The rest of paper is organized as follows. First, we give problem statement and data processing in Section 2. The Framework consisting of off-line classification, on-line detection and pricing regulation. In Section 3, we introduce a Logistic regression method to categorize the detour trips in off-line data. Subsequently, we extend off-line classification to on-line detection via some sophisticated converting in Section 4. Comprehensive pricing analysis and suggestions are described in Section 5. Finally, we conclude our work in Section 6. 

\section{Problem Formulation}

\subsection{Preliminary knowledge}
In this section,  notations  used in this  paper will be first introduced. A road network is represented by a directed graph $G= {\langle N,S \rangle}$, which includes a node set $N$ and a  segment set  $S$. In particular, segment $s$ is a basic unit generated by intersected nodes in $N$, and each segment in a road network owns a specific direction. There are some important concepts defined as follows:  

\textbf{Definition 1}: A \textbf{GPS point} $p$ is denoted by a triple  $\langle lat,lng,t \rangle$, which stands for the latitude, longitude and the GPS generation time of $p$. 

\textbf{Definition 2}: A taxi \textbf{trajectory} $tr$ is a series of GPS points which are generated by an occupied taxi and then ordered by timestamp, 
\begin{center}
$tr=\{p_1,p_2,...,p_m\}$
\end{center}
Obviously, for this trip $p_1$ is the origin position and $p_m$ is the destination position.

Because the GPS information is updated in a couple seconds, one taxi trajectory may contain hundreds (or even thousands) of records, which are mostly redundant in identifying the trip. To reduce the complexity of calculation, we match the taxi trajectory data with a successive segment series of the road network, namely, transforming $tr=\{p_1,p_2,...,p_m\}$ into a series of ordered segments. One of the most popular map matching algorithms is \textit{Hidden Markov Model} which finds the most possible sequence of status given a sequence of observations \cite{newson2009hidden}. Adopting this method, we can transfer a taxi trajectory into an abstract trajectories, which is defined below.

\textbf{Definition 3}: An \textbf{abstract trajectory} $atr$ is a series of segments that are generated by the map matching process.
\begin{center}
$atr=\{ \langle s_1, t_1\rangle,...,\langle s_i, t_i\rangle,...,\langle s_n, t_n\rangle\}$
\end{center}
where $s_i$ indicates the $ith$ segment unit in the trajectory and $t_i$ represents the corresponding timestamp on $s_i$.

To provide E-hailing drivers with real-time traffic guidance, the online ride-hailing platforms will automatically update the route recommendations at each segment of trajectory. The objective of these recommendations ranges from minimizing the total distance to minimizing the travel time of the route, for simplicity, we can set a linear weighted target to achieve a dynamic shortest path and define a function $r(s_i,s_n,t_i)$ to acquire the shortest path:

\begin{equation}
\begin{aligned}
 r(s_i,s_n,t_i) = argmin_{path(s_i,s_n)} [w_1 \times Dis(path(s_i,s_n))+\\
    w_2 \times Et(path(s_i,s_n),t_i)]
\end{aligned}
\label{Eqa:RP_objective}
\end{equation}
where $Dis(path)$ indicates the function calculating the distance of the given path. In addition, $Et(path,t)$ represents the function calculating the estimated travel time of the given path under the time interval $t$. In particular, we just simply assume that we have sufficient taxi trajectories to estimate the travel time among a specific route, after that we can employ the Wide-Deep-Recurrent(WDR) learning model\cite{wang2018learning} to solve it.

There exists a variety of sophisticated and reliable methods to acquire a high-quality route recommendation. Since the main purpose of this paper focus on the trajectory detection, we will establish route plans according to the methods $\footnote{$w_1$ and $w_2$ can be set according to the route plan service of DiDi map}$ proposed in \cite{bast2016route,martin1993systems}. By monitoring the changes among each route plan, the managers of the platforms can get valid evidence on the behaviors of drivers. Therefore, the route plan is defined below.

\textbf{Definition 4}: Given an $atr$, a \textbf{route plan set} $R$ is a set of route recommendations provided for this trip. In other words, each abstract trajectory ($atr$) has a route plan set (route plan).

\begin{center}
$R=\{r(s_1,s_n,t_1),...,r(s_i,s_n,t_i),...r(s_n,s_n,t_n)\}$
\end{center}

After assigning the GPS points to segments and introducing the route plans, we will handle with $atr$ and route plan $R$ in the rest of this paper. Our objectives are to detect the anomalous trajectories in real-time and discover the primary cause of these misbehavior. Formally, the problem is defined as below.

\textbf{Problem}: Given a trajectory $atr$ and its corresponding route plans $R$, several problems in this paper will be tackled includes:

\begin{itemize}
  
  \item
    \textbf{Off-line classification}: Based on the off-line data-set, we should achieve a high detection accuracy for classifying the detour trips and normal trips.
  \item
    \textbf{On-line detection}: With the real-time updating of trajectory and route plans, we should further provide a real-time indicator to monitor the on-going trajectory.
   \item
    \textbf{Long-term regulation}: From the perspective of long-term, we should discover the primary causes of detour behaviors and offer the platforms some suggestions to restrain the happening of detour.
\end{itemize}

Complete notations which will be used in the subsequent analysis are listed in Table \ref{Table:Notation}.

\begin{table}[]
\centering
\caption{Notations List}
\begin{tabular}{p{2cm}|p{6cm}}
\toprule
\textbf{Variable} & \textbf{Explanation }\\[3pt] \midrule
$s$ & Road segment \\[2pt] \midrule
$p$ & GPS point, $p=\langle lat,lng,timestamp \rangle$ \\ [2pt] \midrule
$tr$ & Trajectory, $tr=\{p_1,p_2,...,p_m\}$ \\[2pt] \midrule
\multirow{2}{*}{$atr$} & Abstract trajectory\\
& $atr=\{ \langle s_1, t_1\rangle,...,\langle s_i, t_i\rangle,...,\langle s_n, t_n\rangle\}$ \\[2pt] \midrule
\multirow{2}{*}{$R$} & Route plan of given trajectory $atr$\\
& $R=\{r(s_1,s_n,t_1),...,r(s_i,s_n,t_i),...r(s_n,s_n,t_n)\}$ \\[2pt] \midrule
$\epsilon$ & Destination changing probability\\[2pt] \midrule
$A$ & Historical trajectory data-set, $\forall atr \in A$ \\[2pt] \midrule
$B$ & Historical driver data-set\\[2pt] \midrule
$U$ & Detour utility per time unit of E-hailing drivers \\[2pt] \midrule
$\Theta$ & Indicator of log-odds in categorizing model \\[2pt] \midrule
$\Theta_i$ & Dynamic indicator of log-odds traveling on $\langle s_i, t_i\rangle$ \\[2pt] \midrule
$X^{(1)},X^{(2)}$ & Ratio of detour distance; Ratio of delay time  \\[2pt] \midrule
$X_i^{(1)},X_i^{(2)}$ & Dynamic distance scores and travel time scores traveling on $\langle s_i, t_i\rangle$\\[2pt] \midrule
$w_1,w_2$ & Linear weight of distance and travel time \\[2pt] \midrule
$K,K_0$ & Actual destination, initial destination of trajectory $atr$  \\[2pt] \midrule
$\alpha_1,\alpha_2$ & Fare rate per unit distance and per unit time \\[2pt] \midrule
$\alpha_3,\alpha_4$ & Operating cost of unit distance, opportunity cost of unit time \\[2pt] \midrule
$K,K_0$ & Actual destination, initial destination of trajectory $atr$  \\[2pt] \midrule
$f_0,\mu,\tau$ & Base fare, begin-charged distance and begin-charged time \\[2pt] \midrule
$\beta_0,\beta_1,\beta_2$ & Coefficients to be estimated in categorizing model \\[2pt] \midrule
$\widehat{\beta}_0,\widehat{\beta}_1,\widehat{\beta}_2$ & Estimated coefficients from categorizing model \\[2pt] 

\bottomrule
\end{tabular}
\label{Table:Notation}
\end{table}

\begin{figure}[htbp]
\centering
\includegraphics[width=3.4in]{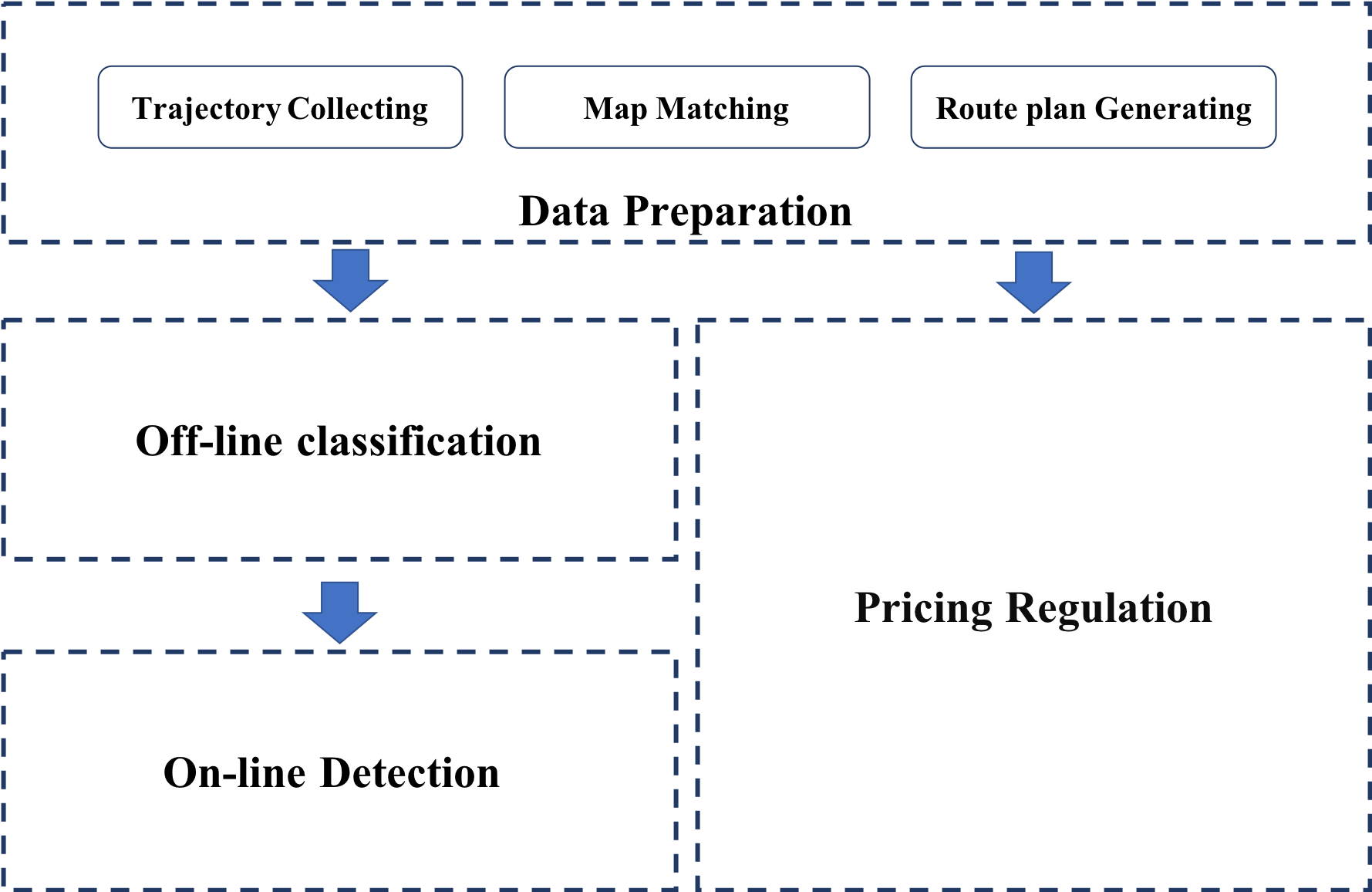}
\caption{Overview of our method}
\label{Fig:Framework}
\end{figure}

Having defined the necessary notations and stated the problem, the proposed framework consists of three phases shown in Figure \ref{Fig:Framework}. Firstly, the result of data preparation will be fed into the off-line classification phase to accurately discover detour trips.
After developing and training an off-line model, we further extend the methods to on-line detection. Parallel to detection mechanism, a pricing regulations have also been introduced in the third phases to eliminate the detour behaviors on E-hailing platforms.

\subsection{Data preparation}

Given a large amount of taxi trajectories consisting of mass GPS points, valid taxi OD pairs can be extracted. Firstly, we can collect the data-set of GPS points for target cities. Secondly, applying the map matching method in \cite{newson2009hidden}, we can transform the GPS points into a series of segments and obtain abstract trajectories $atr$. Meanwhile, we can generate corresponding route plans by introducing the methods in \cite{bast2016route}. Consequently, we obtain a structured mapped indexing table that contains $atr$ identifier, the abstract trajectory $atr$ and the route plans $R$.

In reality, however, we should note that there exists some trips changing their initial destinations without advanced reporting in E-hailing platforms. In order to clear up the purposes of above scenarios, we introduce an destination changing probability defined as follows:

\begin{equation}
\epsilon=1-\frac{Euc(K,K_{0})}{Dis(atr)}
\label{Eqa:epsilon}
\end{equation}
where $Euc(p_i,p_j)$ represents the function acquiring the Euclidean distance between $p_i$ and $p_j$. Meanwhile, $K$, $K_0$ are the actual destination and recorded destination in E-hailing platform, respectively.

In this paper, the probability threshold $\bar{\epsilon}$ is empirically introduced, which aims to identify the purposes of drivers' deviations. If $\epsilon < \bar{\epsilon}$, we think the driver wants to go to an entirely different destination $K$, so we don’t research these special cases in this paper. Nevertheless, if $\bar{\epsilon} \leq \epsilon \leq 1$,we consider difference between $K$ and $K_0$ is caused by accessibility (parking, traffic rules and so on) of the destination $D$ rather than changing destinations.

Consequently, we randomly select December 1th to 31th, 2018 and then choose floating-car data-sets of these days from four major cities (Beijing, Shanghai, Guangzhou, Shenzhen). After removing the abnormal cases with very short travel time (less than 60s), extremely high travel speed (exceeding 120km/h) and absolutely different destinations ($\epsilon < \bar{\epsilon}$), we obtain about millions of samples and thousands of drivers in data-set $A$ and $B$, respectively.

Combining with passengers’ feed-backs and manual verification (processed by customer response system in DiDi Chuxing), all the trips can be labeled as detour or not. To sum up, Table \ref{Table:data_statistics} lists the statistics of these four data-sets. Specifically, the parameter  $\bar{\epsilon}$ is an empirical value set to 0.01 according to the suggestions of experienced map experts from DiDi Chuxing.

\begin{table}[]
\centering
\caption{Statistics of data-set}
\begin{tabular}{p{2.5cm}|p{1cm}p{1cm}p{1cm}p{1cm}}
\toprule
City & Beijing & Shanghai & Guangzhou & Shenzhen \\[2pt]\midrule 
$|A|$ (million) & $>1$ & $>0.5$ & $>0.5$ & $>0.5$ \\[2pt] 
$|B|$ (thousand) & $>30$ & $>30$ & $>30$ & $>30$ \\[2pt] 
Labeled $|A|$ & $41227$ & $27236$ & $17747$ & $21281$ \\[2pt] 
Period & \multicolumn{4}{c}{2018/12/01-2018/12/31}\\[2pt]
Training(\%) & \multicolumn{4}{c}{40}\\[2pt]
Testing (\%) & \multicolumn{4}{c}{60}\\[2pt]
\bottomrule
\end{tabular}
\label{Table:data_statistics}
\end{table}

\section{Off-line classification}

\subsection{Logistic regression based categorizing}

In order to characterize the anomaly degree of given trajectory, we need to take both the route distance and travel time into consideration, and the principal idea of this method is to compare the difference between the 
actual trajectory $atr$ and the initial route plan $r_1$ starting from the origin $s_1$ in terms of the distance level and the time level. Therefore, the distance-based feature $X^{(1)}$ and the time-based feature $X^{(2)}$ of the given trajectory $atr$ can be defined as:

\begin{equation}
X^{(1)}=\frac{Dis(atr)}{Dis(r(s_1,s_n,t_1))}-1
\label{Eqa:X1_eq}
\end{equation}

\begin{equation}
X^{(2)}=\frac{At(atr)}{Et(r(s_1,s_n,t_1),t_1)}-1
\label{Eqa:X2_eq}
\end{equation}
where $At(atr)$ represents the actual travel time of $atr$.

Given a trajectory $atr$ and initial route plan $r(s_1,s_n,t_1)$, $X^{(1)}$ can be treated as the ratio of detour distance and $X^{(2)}$ can be treated as ratio of delay time. Hence, we further apply the logistic regression model to categorizing the detour and normal trips. 

Logistic regression (also known as logit model) is widely used to model the outcomes of a categorical dependent variable \cite{czepiel2002maximum}. Logistic regression measures the relationship between the categorical dependent variable and one or more independent variables by estimating probabilities using a logistic function, which is the cumulative logistic distribution. Providing that the detour indicator is a discrete variable $Y_j$ ($Y_j=1$ represents a detour trip) with factors $X_j=(X_j^{(1)},X_j^{(2)})$ as independent variables, an logistic model can be written in terms of cumulative probability of a specific detour indicator for a given set of factors $X_j$:

\begin{equation}
P(Y_j=1|X_j )=\frac{1}{1+e^{\beta_0+\beta_1 \times X_j^{(1)}+ \beta_2 \times X_j^{(2)}}}
\label{Eqa:condition_probability}
\end{equation}

Applying logit transformation to the cumulative probability:

\begin{equation}
ln {\frac{P(Y_j=1|X_j)}{(1-P(Y_j=1|X_j)}}=\beta_0+\beta_1 \times X_j^{(1)}+ \beta_2 \times X_j^{(2)}
\label{Eqa:ln_condition_probability}
\end{equation}
where $\frac{P(Y_j=1|X_j)}{(1-P(Y_j=1|X_j)}$ is called the odds, which represents the ratio of the probability of detour to the probability of non-detour. Similarly, $ln \frac{P(Y_j=1|X_j)}{(1-P(Y_j=1|X_j)}$ is a log-odds, which is denoted as $\Theta_j$ in the rest of paper. 

Having transformed the probability into a log-odds, the logit function linearizes the association between the probability and independent variables. Though as a generalized linear model, simple linear regression estimation methods like least-squares estimation are not applicable. The maximum likelihood estimation (MLE) is applied instead. The log-transformation is widely used to deal with the likelihood function in practice. Therefore, if we set the $P(Y_j=1|X_j)=h(X_j)$, then we have a log-likelihood function:
\begin{equation}
L(\beta)=\sum_{n=1}^N {[Y_j \times ln(h(X_j))+ (1-Y_j)\times ln(1-h(X_j))]}
\end{equation}

The MLE method is thus deduced as:
\begin{equation}
\widehat{\beta}=argmax \quad L(\beta)
\label{Eqa:MLE}
\end{equation}

Common optimization techniques are used in order to solve the MLE method involve \textit{Gradient Descent Algorithm} and \textit{Quasi-Newton Methods}. In this paper, for brevity, optimization algorithm is not provided. Alternatively, we can solve the above model by function \textit{LogisticRegression} of Scikit-learn package from Python.

\subsection{Model evaluation}

\begin{figure}[htbp]
\centering
\includegraphics[width=3.4in]{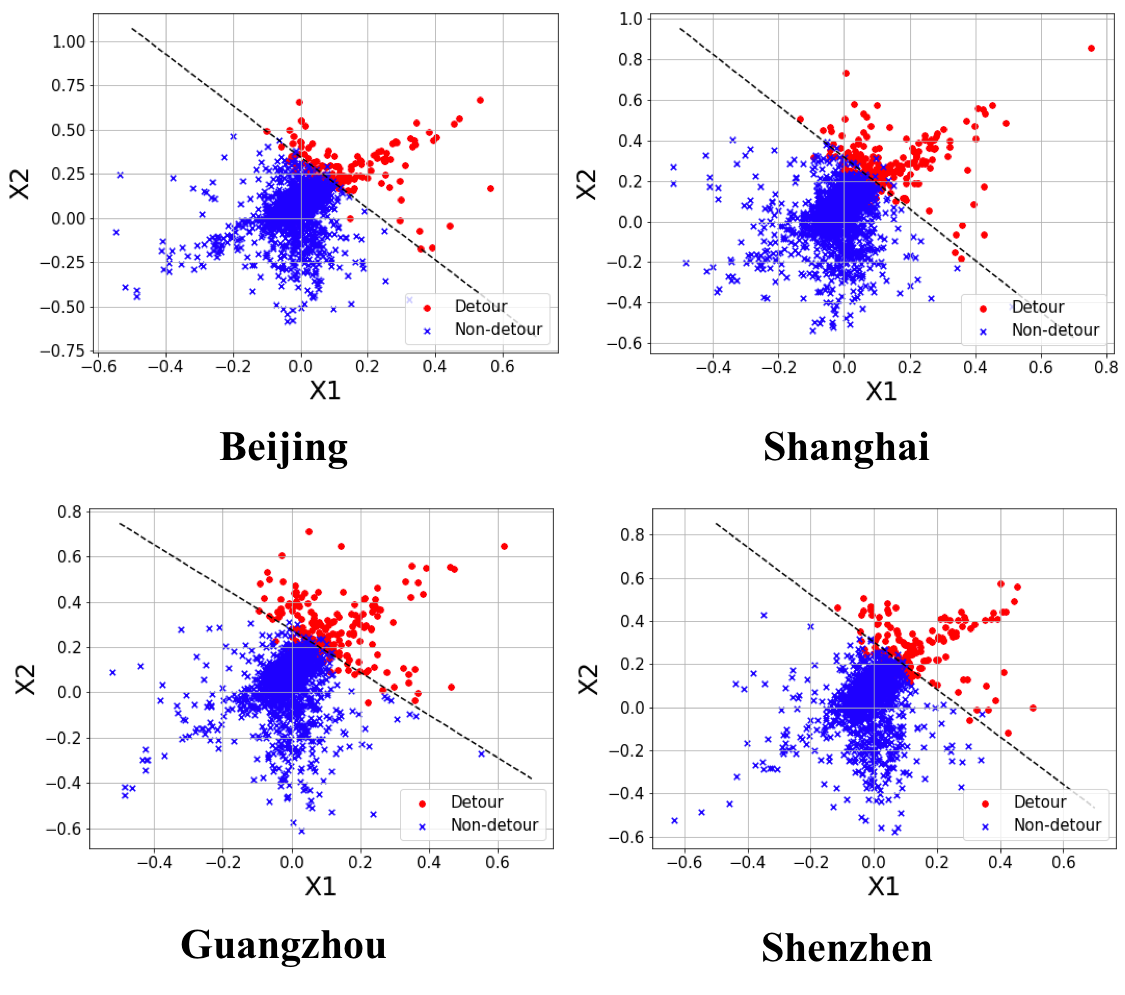}
\caption{The Classification in four cities}
\label{Fig:Classfication}
\end{figure}

The results of the logistic regression model are presented in Table \ref{Table:LR_result}. The results show that the coefficients of  $X^{(1)}$ and $X^{(2)}$ differ across different cities, but signs of the coefficients remain positive, and the p-values show that both distance-based variable $X^{(1)}$ and time-based variable $X^{(2)}$ are significant. The aforementioned analysis is consistent with the reality. Moreover, the classifying boundaries of four cities are depicted in Figure \ref{Fig:Classfication}.

\begin{table}[]
\centering
\caption{Outputs for the logistic regression model}
\begin{tabular}{p{1.5cm}|p{1cm}|p{1cm}p{1cm}p{1cm}p{1cm}}

\toprule
City & Variable & Coefficient & Standard Error & Chi-Square&P-value \\[2pt]\midrule 

\multirow{3}{2cm}{Beijing} & Intercept & $-8.8620$ & $0.725$ & $-12.218$ & $0.000$ \\[2pt] 
& $X^{(1)}$ & $41.5258$ & $3.906$ & $10.632$ & $0.002$ \\[2pt]
& $X^{(2)}$ & $28.5575$ & $2.643$ & $10.804$ & $0.000$ \\[2pt] \midrule

\multirow{3}{2cm}{Shanghai} & Intercept & $-8.6631$ & $0.577$ & $-15.010$ & $0.000$ \\[2pt] 
& $X^{(1)}$ & $38.7750$ & $3.129$ & $12.391$ & $0.000$ \\[2pt]
& $X^{(2)}$ & $30.4732$ & $2.272$ & $13.415$ & $<0.001$ \\[2pt]
\midrule

\multirow{3}{2cm}{Guangzhou} & Intercept & $-10.0029$ & $0.803$ & $-12.450$ & $0.000$ \\[2pt] 
& $X^{(1)}$ & $37.4299$ & $3.350$ & $11.175$ & $0.000$ \\[2pt]
& $X^{(2)}$ & $39.8327$ & $3.471$ & $11.476$ & $0.000$ \\[2pt]
\midrule

\multirow{3}{2cm}{Shenzhen} & Intercept & $-9.3678$ & $0.777$ & $-12.052$ & $0.000$ \\[2pt] 
& $X^{(1)}$ & $37.6850$ & $3.715$ & $10.144$ & $0.000$ \\[2pt]
& $X^{(2)}$ & $34.2923$ & $3.257$ & $10.529$ & $0.000$ \\[2pt]
\bottomrule

\end{tabular}
\label{Table:LR_result}
\end{table}

\begin{figure}[htbp]
\centering
\includegraphics[width=3.4in]{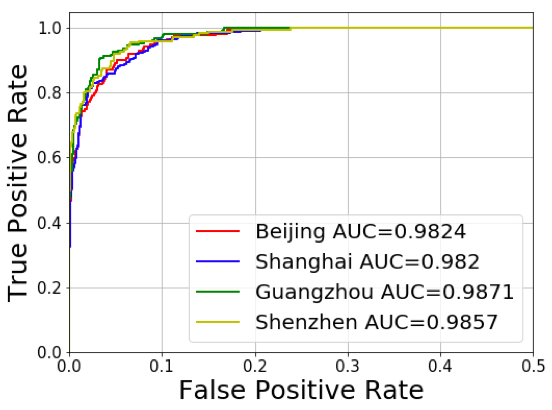}
\caption{The ROC curves of the Logistic regression model}
\label{Fig:ROC}
\end{figure}

The evaluation for the model in this paper is using the $AUC$ (Area Under ROC Curve). In practice, true positive rate $TPR$ (the fraction of anomalous data that is successfully detected) and false positive rate $FPR$ (the fraction of normal ones that is predicted to be anomalous) are two important measures to evaluate the performance of an anomaly detection method. Obviously, a good anomaly detection method should have both high $TPR$ and low $FPR$. The $ROC$ curve shows the $TPR$ (y-axis) against the $FPR$ (x-axis), and the $AUC$ value is defined as the area under the $ROC$ curve. Figure \ref{Fig:ROC} depicts the $ROC$ curves of proposed method on four datasets in Table \ref{Table:iBAT}. We can find that the proposed method is able to achieve high detection rate whilst keeping low false alarm rate. For all data-sets, over 90\% of detour trajectories can be detected at a 10\% false alarm rate.

Without loss of generality, a typical comparison analysis is also conducted, we compare the $AUC$ value of LR with the iBAT method \cite{zhang2011ibat} as shown in Table \ref{Table:iBAT}, we can see that LR achieves quite high $AUC$ values ($>0.98$ on all datasets) and the iBAT method achieves lower $AUC$ values ($<0.94$ on all datasets), suggesting that the logistic method under our framework outperforms the iBAT method to detect the detour behaviors. This is due to the fact that iBAT suffers from the problem of data sparsity, if a trajectory is infrequent, it’s difficult to detect outlier based on the theory of similarity. While our method can overcome this obstacle to achieve a remarkable performance by combining the corresponding route plan information.

\begin{table}[]
\centering
\caption{The AUC value of LR and iBAT method}
\begin{tabular}{p{1.5cm}|p{1cm}p{1cm}p{1cm}p{1cm}}
\toprule
Methods & Beijing & Shanghai & Guangzhou & Shenzhen \\[2pt]\midrule 
LR & $0.9824$ & $0.9820$ & $0.9871$ & $0.9857$ \\[2pt] 
iBAT & $0.9327$ & $0.9228$ & $0.8935$ & $0.9324$ \\[2pt] 
\bottomrule
\end{tabular}
\label{Table:iBAT}
\end{table}

\section{On-line detection}

In this phases, we will further extend off-line classification to on-line detection. According to the on-line setting, the trajectory and route plan will be updated at every timestamp along with the movement of the trip. Therefore, we need to provide a judgement at each timestamp to identify whether this trip is anomalous or not.

\subsection{Dynamic scoring}

Given the unfinished trajectory $atr(s_1,s_i)$ traveling on $\langle s_i, t_i\rangle$, we further introduce two indicators $X_i^{(1)}$ and  $X_i^{(2)}$ to achieve dynamic scoring:

\begin{equation}
X_i^{(1)}=\frac{Dis(atr(s_1,s_i))+Dis(r(s_i,s_n,t_i))}{Dis(r(s_1,s_n,t_1))}-1
\label{Eqa:X1_eq_online}
\end{equation}

\begin{equation}
X_i^{(2)}=\frac{At(atr(s_1,s_i))+Et(r(s_i,s_n,t_i),t_i)}{Et(r(s_1,s_n,t_1),t_1)}-1
\label{Eqa:X2_eq_online}
\end{equation}
where $Dis(atr(s_1,s_i))+Dis(r(s_i,s_n,t_i))$ in equation \ref{Eqa:X1_eq_online} is the the estimated total distance
when reaching $\langle s_i, t_i\rangle$. Similarly, $X_i^{(2)}$ in equation \ref{Eqa:X2_eq_online} indicates the estimated travel time when reaching $\langle s_i, t_i\rangle$. Therefore, the distance-based feature $X_i^{(1)}$ and the time-based feature $X_i^{(2)}$ of given $atr(s_1,s_i)$ can be derived.

\begin{figure}[htbp]
\centering
\includegraphics[width=3.4in]{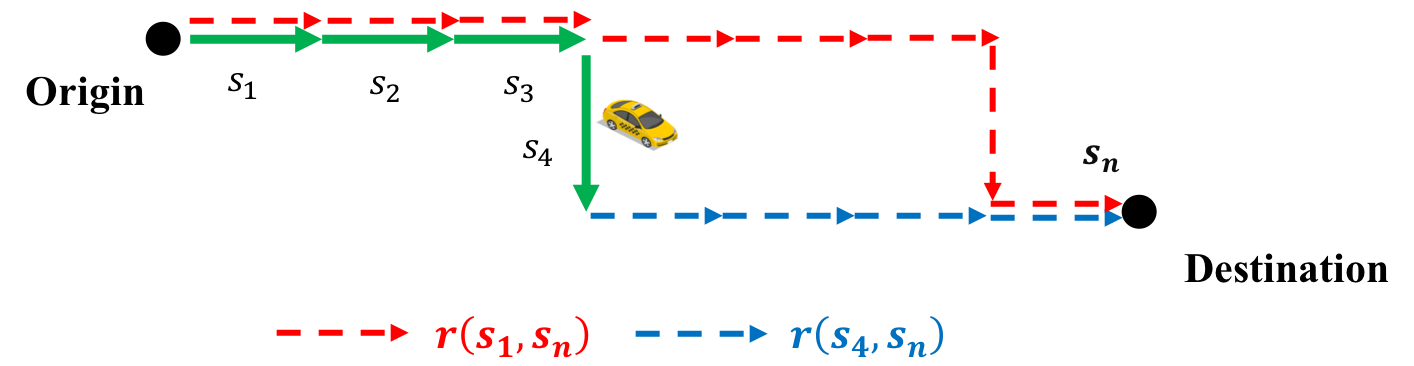}
\caption{Examples of dynamic scoring}
\label{Fig:Dynamic_scoring}
\end{figure}

For a better understanding, Figure \ref{Fig:Dynamic_scoring} presents an unfinished trajectory $atr(s_1,s_4)$ traveling on $\langle s_3, t_4\rangle$. Intuitively, $X_4^{(1)}$, $X_4^{(2)}$ can be derived from the distance difference and travel time difference between red sequences and green merging blue sequences, respectively. In practice, since there will exist a significant difference between the estimated total distance and the initially recommended distance if deviation occurs, the advantage of using the measurements in Equation \ref{Eqa:X1_eq_online} and \ref{Eqa:X2_eq_online} contributes to exactly identifying the deviation behaviors such as avoiding congestion, experienced choices and so on. We can easily observe three kinds of scenarios of $X_i^{(1)}$ and $X_i^{(2)}$ as follows:

\begin{itemize}
  \item 
	If $X_i^{(1)}>0$ and $X_i^{(2)}>0$, we can observe the deviation occurring in $s_i$ leads to a worse condition, which means more travel distance and more time-consuming.
  \item 
	if $X_i^{(1)}>0$ or $X_i^{(2)}>0$, the deviation occurring in $s_i$ results in complex situations and needs further discussions.
	\begin{itemize}
        \item [a)]
        The driver chooses another longer route to avoid the traffic jam compared to the initial route plan ($X_i^{(1)}>0$ and $X_i^{(2)}<0$).
        \item [b)]
        The driver thinks he takes a shortcut compared to the original plan, but unfortunately, he cuts off in traffic ($X_i^{(1)}<0$ and $X_i^{(2)}>0$).
    \end{itemize}
  \item
    if $X_i^{(1)}<0$ and $X_i^{(2)}<0$, it’s obvious the driver takes a better route and we need to optimize our recommendation strategy.
\end{itemize}

After computing the ongoing scores $X_i^{(1)}$ and $X_i^{(2)}$, we can calculate log-odds $\Theta_i$ at current segment $\langle s_i, t_i\rangle$:
\begin{equation}
\Theta_i=\widehat{\beta}_0+\widehat{\beta}_1 \times {X_i}^{(1)}+ \widehat{\beta}_2 \times {X_i}^{(2)}
\label{Eqa:Theta_i}
\end{equation}
where $\widehat{\beta}_0$, $\widehat{\beta}_1$ and $\widehat{\beta}_2$ can be estimated from off-line categorizing model from equation \ref{Eqa:MLE}.

As a consequence, the log-odds $\Theta_i$ should be treated as a significant indicator in on-line framework to detect the driver's behavior:

\begin{itemize}
  \item If $\Theta_i >0$, it's obvious that the on-going trip has fall into the category of detour. Therefore, a warning message will be triggered to remind drivers to regulate their behaviors.
  \item Otherwise, the on-going trip will be treated as a normal case. It should note that if a warning message has been activated in previous segment, the platform will cancel the warning and inform the drivers. 
\end{itemize}

In E-hailing companies, the analysis of taxi driver behavior based on our online detection method could provide an efficient evaluation of the driver’s performance. Meanwhile, our online detection method could help E-hailing companies build a good service environment with fine, self-disciplined taxi drivers.

\subsection{Experimental performance}

\begin{figure}[htbp]
\centering
\includegraphics[width=3.4in]{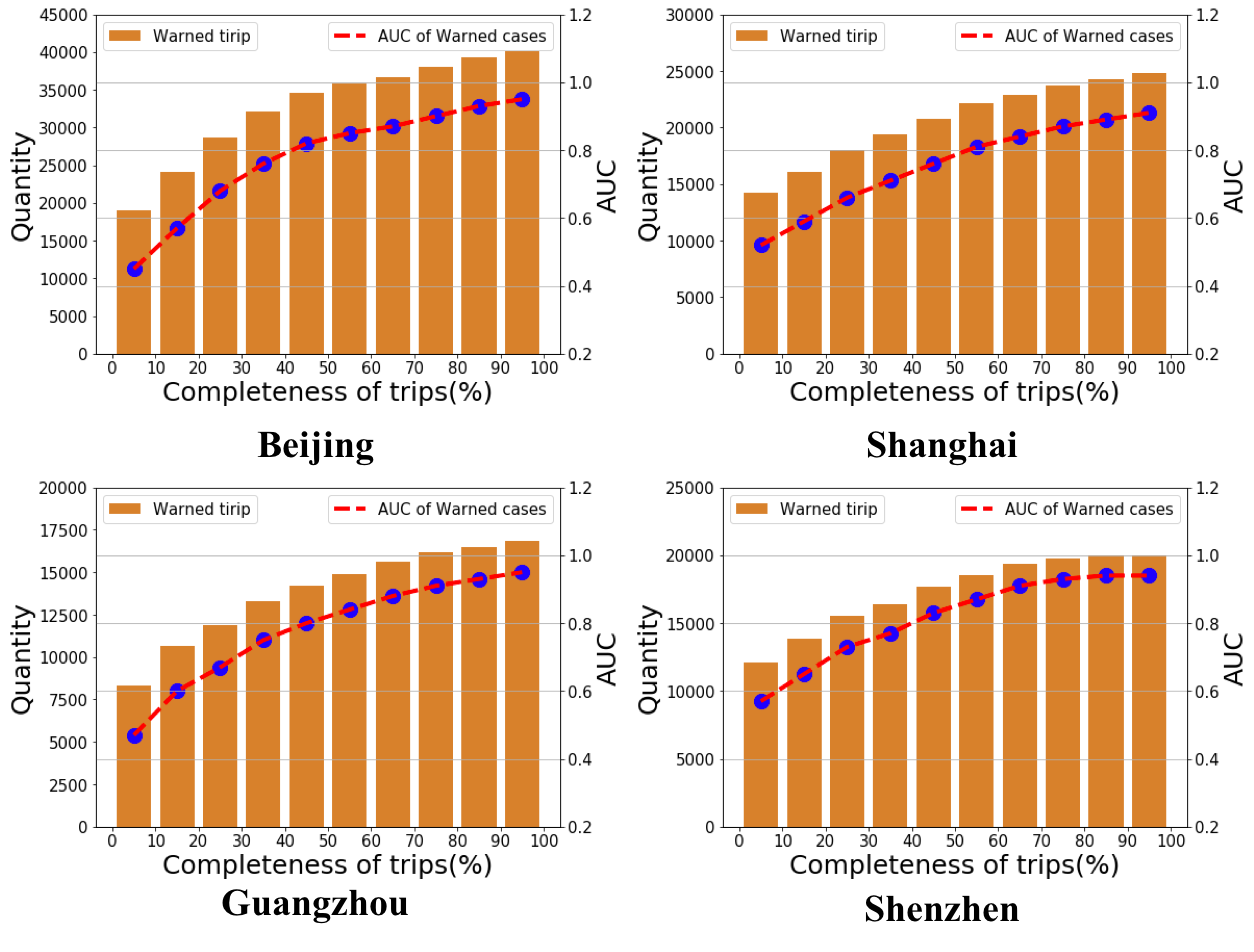}
\caption{The AUC of warned trips of online detection in four cities}
\label{Figure:Warning_AUC}
\end{figure}

To evaluate the effectiveness of the on-line detection phase, we divided a trip into ten stages in terms of its completeness. Based on the statistics of warned trips quantity among each stage, we plot the AUC of each stage in the Figure \ref{Figure:Warning_AUC}. There are two interesting discoveries in the figures: Firstly, the warned trips quantity dramatically increase among the first 50\% stages, in other words, the majority of detour behaviors may happen in the first half of the trip. Secondly, due to the limited information known in the first 50\% stages, AUCs are not very high (less than 0.8) and there will exists lots of misjudged warnings. However, with going of the trip, the majority of misjudged warnings will be cancelled in our on-line mechanism and ultimately AUCs will achieve an excellent performance($AUC>0.9$) at the tail stages.

\section{Long-term pricing regulation}

\subsection{Pricing analysis}

From perspective of long-term, it is necessary to note that pursuing higher income might be the primary cause of detour trips. Therefore, rather than on-line detection, regulating proper fare rate will exert a significant influence on the drivers' income level, further eliminate the number of detours in a long-term. In order to detect these deliberate misbehavior related to drivers' income, we may discuss the pricing mechanism in E-hailing platform at first, a price-wise linear fare structure has been widely used in E-hailing taxi industries as below:

\begin{equation}
f(atr) = f_0+\alpha_1 \times \widetilde{D}(atr,\mu) + \alpha_2 \times \widetilde{T}(atr,\tau)
\label{Eqa:price_eq}
\end{equation}
where $f_0$ denotes the base fare if the total distance and the overall travel time do not exceed $\mu$ and $\tau$. Moreover, $\alpha_1$ and $\alpha_2$ represent the fare rate per unit distance and per unit time, respectively. Specifically, here we introduce $\widetilde{D}(atr,\mu)$ and $\widetilde{T}(atr,\tau)$ to calculate the extra distance and travel time based on $\mu$ and $\tau$, respectively.

\begin{equation}
\widetilde{D}(atr,\mu) = H(Dis(atr)-\mu) \times [Dis(atr)-\mu]
\label{Eqa:over_dis_eq}
\end{equation}

\begin{equation}
\widetilde{T}(atr,\tau)= H(At(atr)-\tau) \times [At(atr)-\tau]
\label{Eqa:over_time_eq}
\end{equation}

In Equation \ref{Eqa:over_dis_eq} and \ref{Eqa:over_time_eq}, the Heaviside function $H(\cdot)$ is established to verify whether the variable is over zero:

\begin{equation}
H(x)=\left\{
\begin{array}{ccc}
0       &      & {if\; x<=0}\\
1      &      & {otherwise}
\end{array} \right. 
\label{Eqa:Heaviside}
\end{equation}

To conduct a deep analysis upon anomaly related to income, we define a detour utility (per time unit) $U$ to gain some basic understandings of the detour behaviors:

\begin{equation}
U=\alpha_1 \times v +\alpha_2 - \alpha_3 \times v -\alpha_4
\label{Eqa:utility_minute}
\end{equation}
where $\alpha_3$ represents the coefficient of operating costs per unit distance, and $\alpha_4$ indicates the loss of opportunity cost per unit time, which depends on the base fare $f_0$. In addition, $v$ indicates the average serving speed of the drivers. Specifically, $v$ can just be scaled to distance unit per time unit and estimated in Table \ref{Table:Pricing_parameter}.

According to the utility function $U$, we can have a quantified indicator of a detour behavior in term of its monetary revenue. If the driver takes a detour in a time unit, he will receive a monetary income $\alpha_1 \times v +\alpha_2$, while pay for a fuel cost $\alpha_3 \times v $ and opportunity cost $\alpha_4$. As a consequence, the changes of detour utility will have a significant influence on driver's intention upon detour or not.

\begin{table}[]
\centering
\caption{The pricing regulation in DiDi Chuxing}
\begin{tabular}{p{0.5cm}|p{0.8cm}p{0.8cm}p{0.8cm}p{0.8cm}p{0.8cm}p{0.8cm}}
\toprule
&City& 00:00-06:00 & 06:00-12:00 & 12:00-17:00 & 17:00-21:00 & 21:00-24:00 \\[2pt]\midrule 
\multirow{4}{2cm}{$\alpha_1$}& BJ & $2.15$ & $1.80$ & $1.45$ & $1.50$ & $2.15$ \\[2pt]
& SH & $3.20$ & $2.30$ & $2.30$ & $2.30$ & $3.20$ \\[2pt]
& GZ & $2.60$ & $2.50$ & $1.90$ & $2.30$ & $2.60$ \\[2pt]
& SZ & $2.90$ & $2.05$ & $2.05$ & $2.30$ & $2.95$ \\[2pt]
\midrule
\multirow{4}{2cm}{$\alpha_2$}& BJ& $0.80$& $0.80$ & $0.40$ & $0.80$ & $0.80$ \\[2pt]
& SH& $0.60$& $0.70$ & $0.60$ & $0.70$ & $0.60$ \\[2pt]
& GZ& $0.40$& $0.40$ & $0.30$ & $0.40$ & $0.40$ \\[2pt]
& SZ& $0.55$& $0.65$ & $0.55$ & $0.65$ & $0.55$ \\[2pt]
\midrule

\multirow{4}{2cm}{$v$}& BJ& $0.629$& $0.467$ & $0.449$ & $0.435$ & $0.526$ \\[2pt]
& SH& $0.581$& $0.382$ & $0.418$ & $0.359$ & $0.483$ \\[2pt]
& GZ& $0.612$& $0.442$ & $0.415$ & $0.409$ & $0.478$ \\[2pt]
& SZ& $0.593$& $0.397$ & $0.373$ & $0.403$ & $0.457$ \\[2pt]
\midrule

\multirow{4}{2cm}{$f_0$}&BJ &\multicolumn{5}{c}{$13\;Yuan$}\\[2pt]
&SH &\multicolumn{5}{c}{$14\;Yuan$}\\[2pt]
&GZ &\multicolumn{5}{c}{$11\;Yuan$}\\[2pt]
&SZ &\multicolumn{5}{c}{$12\;Yuan$}\\[2pt]
\midrule

\multirow{4}{2cm}{$\mu$}&BJ &\multicolumn{5}{c}{$3\;Km$}\\[2pt]
&SH &\multicolumn{5}{c}{$3\;Km$}\\[2pt]
&GZ &\multicolumn{5}{c}{$2\;Km$}\\[2pt]
&SZ &\multicolumn{5}{c}{$3\;Km$}\\[2pt]
\midrule

\multirow{4}{2cm}{$\tau$}&BJ &\multicolumn{5}{c}{$10\;Min$}\\[2pt]
&SH &\multicolumn{5}{c}{$10\;Min$}\\[2pt]
&GZ &\multicolumn{5}{c}{$4\;Min$}\\[2pt]
&SZ &\multicolumn{5}{c}{$8\;Min$}\\[2pt]
\midrule

$\alpha_3$ &\multicolumn{6}{c}{$0.5\;Yuan/Km$} \\[2pt] 
\bottomrule
\end{tabular}
\label{Table:Pricing_parameter}
\end{table}

\subsection{Suggested pricing policies}

In the previous section, we have presented the pricing mechanism and define the utility functions, it's obvious that there exists some relationships between utility and detour behaviors. 

\begin{figure}[htbp]
\centering
\includegraphics[width=3.4in]{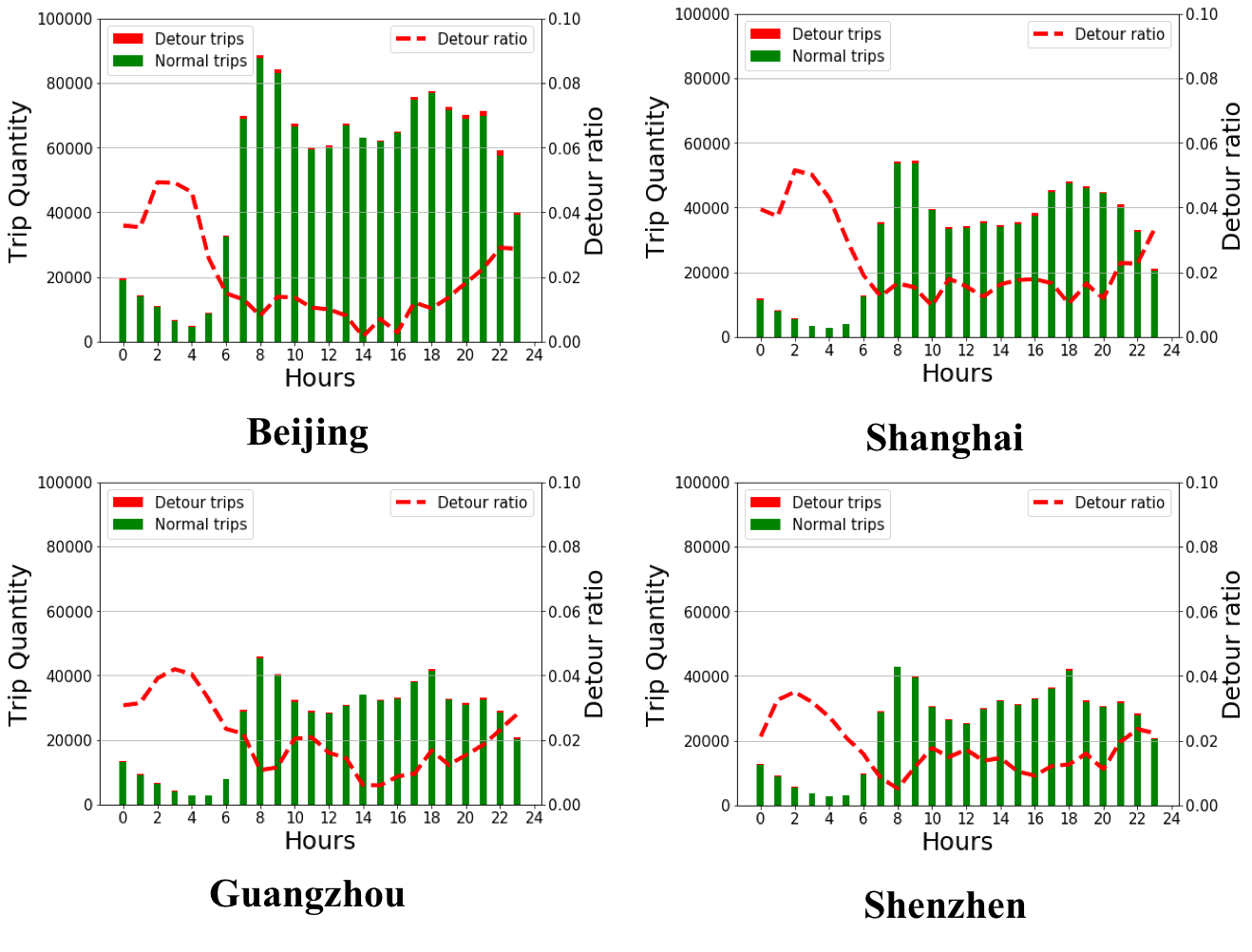}
\caption{Detour Distributions under different time interval in a day}
\label{Fig:Detour_ratio}
\end{figure}

Therefore, we introduce a detour ratio to describe the detour intensity in each time interval, which is the ratio of detour quantity and total quantity. Figure \ref{Fig:Detour_ratio} displays the changes of trip quantity and detour ratio under different time interval in a day. As intuitions suggest, all the trend of order quantity in four cities show the morning-peak trend (among 07:00-10:00),  evening-peak trend (among 17:00-19:00) and the off-peak trend in early hours (among 3:00-5:00). While, as a sharp contrast, the detour ratio climbs at the peak in early hours (among 2:00-5:00) then gradually decreases and tends to some small fluctuations in the daytime. 

\begin{figure}[htbp]
\centering
\includegraphics[width=3.4in]{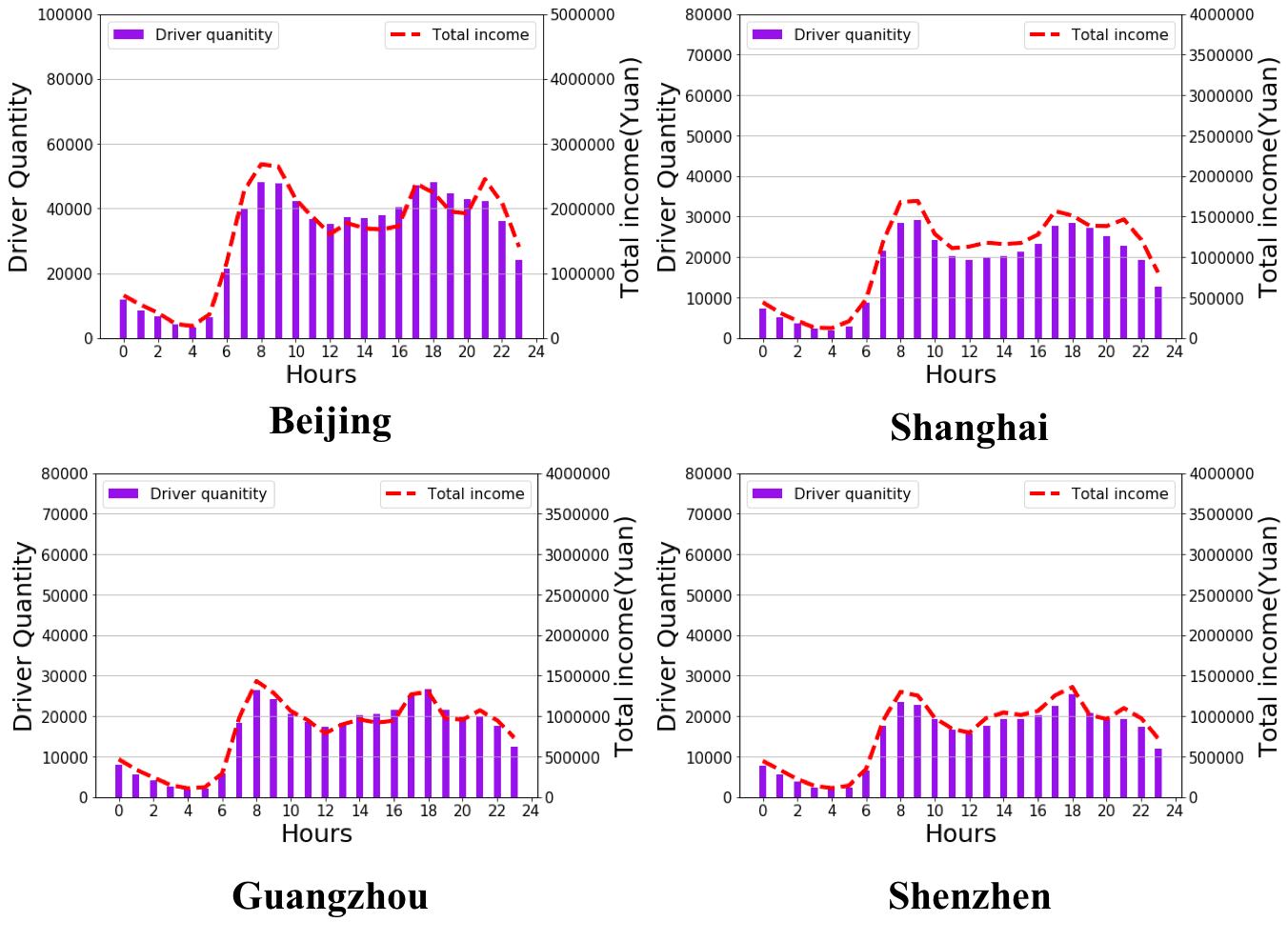}
\caption{Total income and driver quantity in four cities}
\label{Fig:Income_driver}
\end{figure}

To explain the changes of detour ratio, we further investigate the total income of E-hailing platforms and on-duty drivers quantity in a day. As intuitively presented in Figure \ref{Fig:Income_driver}, the total income arrives at the high value during the morning-peak and the evening-peak. However, due to a lower demand quantity and fewer drivers, it drops sharply in early hours (among 2:00-5:00). Meanwhile, Table \ref{Table:Pricing_parameter} list other pricing parameters derived from DiDi Chuxing, which are also public on the DiDi's mobile APP $\footnote{$\alpha_3$ can be obtained from 2018 December China oil prices statistics}$ . In addition, we can figure out $\alpha_4$ as the average income per time unit (minute) of all drivers from aforementioned statistic:

\begin{equation}
\alpha_4=\frac{\sum_{|A|} f(atr)}{|B| \times 60}
\label{Eqa:alpha_4}
\end{equation}

\begin{figure}[htbp]
\centering
\includegraphics[width=3.4in]{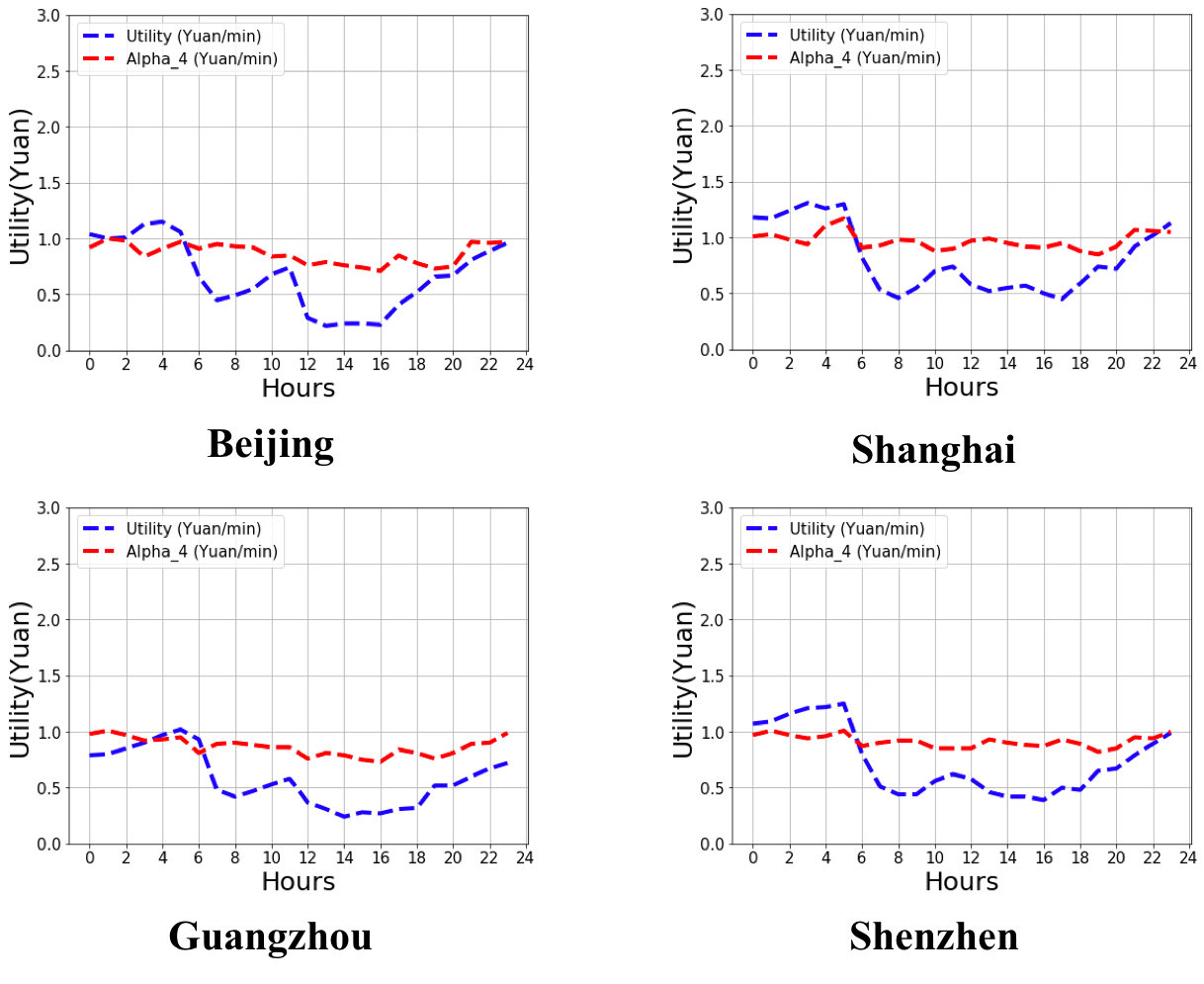}
\caption{Utilities in four cities}
\label{Fig:Utility_alpha4}
\end{figure}

As we have mentioned above, each driver will have a detour utility from equation \ref{Eqa:utility_minute}, in this experiment we particularly focus on the driver's utility per minute in each time interval, thus, we can provide a statistic of detour utility per minute in Figure \ref{Fig:Utility_alpha4}. To evaluate the gains of detour behaviors, the $\alpha_4$ has been also depicted in Figure \ref{Fig:Utility_alpha4} to give a comparison with detour utility. As suggested in Figure \ref{Fig:Utility_alpha4}, it is obvious that the detour utility falls under $\alpha_4$ during the daytime, in other words, taking a detour during daytime may exert a negative effect on increasing monetary income. Contrast to the daytime, the detour utility arises beyond $\alpha_4$ during early hours and approaches $\alpha_4$ in mid-night, which means that deliberate detour could bring much more profit than normal driving. 

\begin{figure}[htbp]
\centering
\includegraphics[width=3.4in]{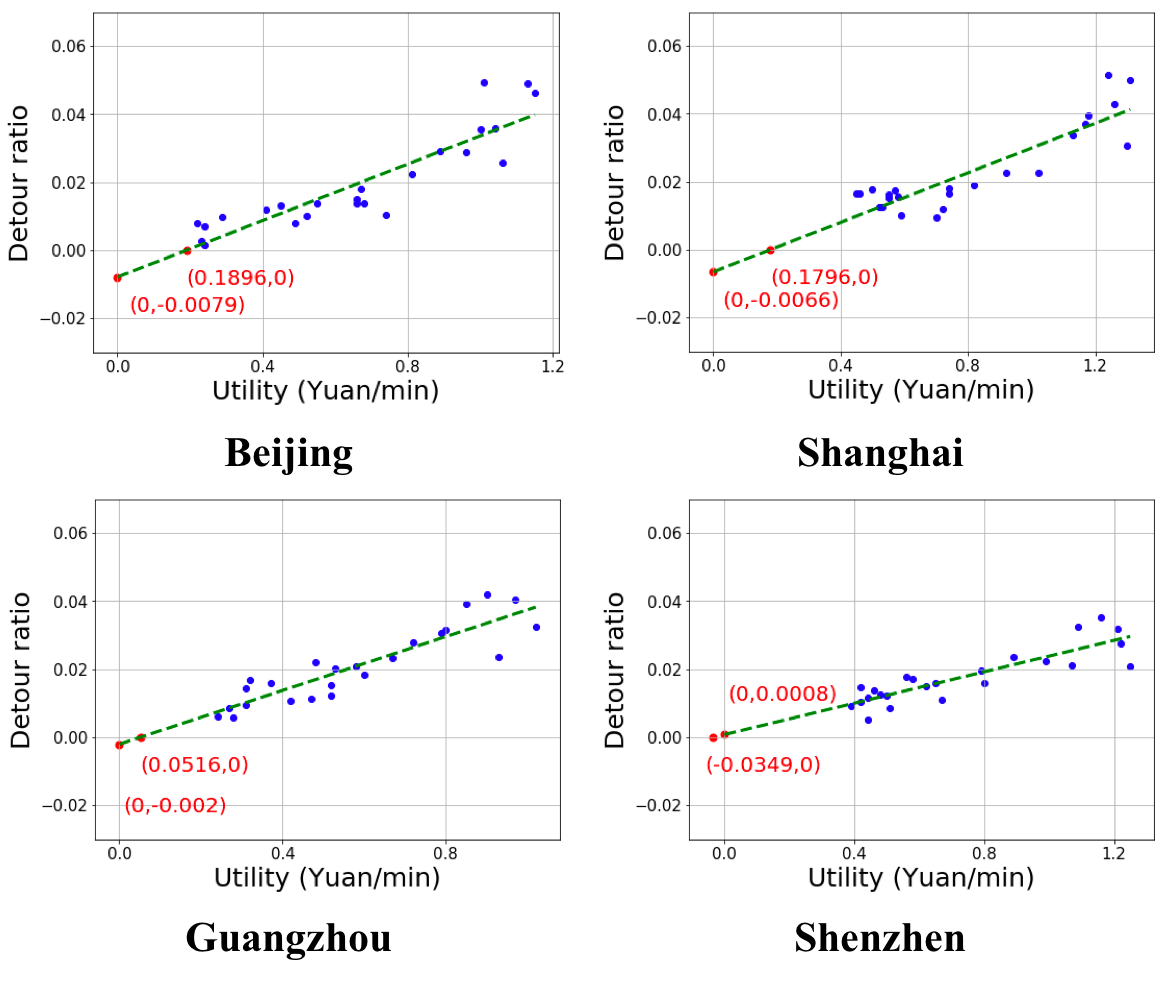}
\caption{Relationship between detour ratio and Utility}
\label{Fig:linear_regression}
\end{figure}

To give an intuitive suggestion, linear regression is introduced to model the relationship between detour ratio and detour utility among each time interval. Intuitively, Figure \ref{Fig:linear_regression} shows the approximately linear correlations between two variables and plots the linear function as well. Furthermore, coefficients, detour ratio intercepts, utility intercepts (here we use $U_0$ to represent the utility of utility intercept) and R-squared in Table \ref{Table:linear_regression}. Obviously, each linear model achieve a remarkable $R^2$ over 75\%, which indicates that our models have achieved good fitting.

\begin{table}[]
\centering
\caption{Statistics of linear regression}
\begin{tabular}{p{2cm}|p{1cm}p{1cm}p{1cm}p{1cm}}
\toprule
City & Beijing & Shanghai & Guangzhou & Shenzhen \\[2pt]\midrule 
Coefficient& $0.0415$ & $0.0365$ & $0.0395$ & $0.0230$ \\[2pt] \midrule
Intercept& $-0.0079$ & $-0.0066$ & $-0.0020$ & $0.0008$ \\[2pt] \midrule
$U_0$& $0.1896$ & $0.1796$ & $0.0516$ & $-0.0349$ \\[2pt] \midrule
$R^2$& $0.806$ & $0.778$ & $0.804$ & $0.757$ \\[2pt] 
\bottomrule
\end{tabular}
\label{Table:linear_regression}
\end{table}

Having established the linear relationships between detour ratio and detour utility, according to the intercept of detour utility, $U_0$ can be calculated to represent the utility that all the drivers are unwilling to take detours. As intuition suggests in Figure \ref{Fig:linear_regression}, all the detour utilities in a day are above $U_0$. In order to motivate drivers to take normal trajectory and eliminate detour behaviors as much as possible, we need to guarantee the detour utility coincides with $U_0$ in each time interval. 

To achieve this goal, we need to modify the pricing mechanism mentioned above, here $\Delta \alpha_1$ and $\Delta f_0$ are introduced to denote the adjustment of $\alpha_1$ and $f_0$, respectively. Specifically, $\Delta \alpha_1$ and $\Delta f_0$ in each time interval need to meet the following constraint:

\begin{equation}
\begin{aligned}
(f_0+\Delta f_0)+(\alpha_1+\Delta \alpha_1) \times \overline{D} + \alpha_2 \times \overline{T} =\\ f_0+\alpha_1 \times \overline{D} + \alpha_2 \times \overline{T}
\end{aligned}
\label{Eqa:Pricing_equilibrium}
\end{equation}
where $\overline{D}$, $\overline{T}$ represent the average distance and travel time exceeding $\mu$ and $\tau$. Based on equation $\ref{Eqa:Pricing_equilibrium}$, we can ensure that the average trip price remains unchanged after adjusting $\Delta \alpha_1$ and $\Delta f_0$.

On the other hand, the detour utility in each time interval should satisfy:

\begin{equation}
(\alpha_1+\Delta \alpha_1) \times v +\alpha_2 - \alpha_3 \times v -(\alpha_4+\Delta \alpha_4)=U_0
\label{Eqa:Utility_equilibrium}
\end{equation}

Equation \ref{Eqa:Utility_equilibrium} ensures that the detour utility equals to $U_0$ after adjusting pricing mechanism, and we should note that $\Delta \alpha_4$ is the changes of $\alpha_4$, which is associated with $\Delta f_0$ and can be derived below:

\begin{equation}
\begin{aligned}
\Delta \alpha_4= \frac{|A| \times \Delta f_0}{|B| \times 60}
\end{aligned}
\label{Eqa:Delta_alpha4}
\end{equation}

\begin{figure}[htbp]
\centering
\includegraphics[width=3.4in]{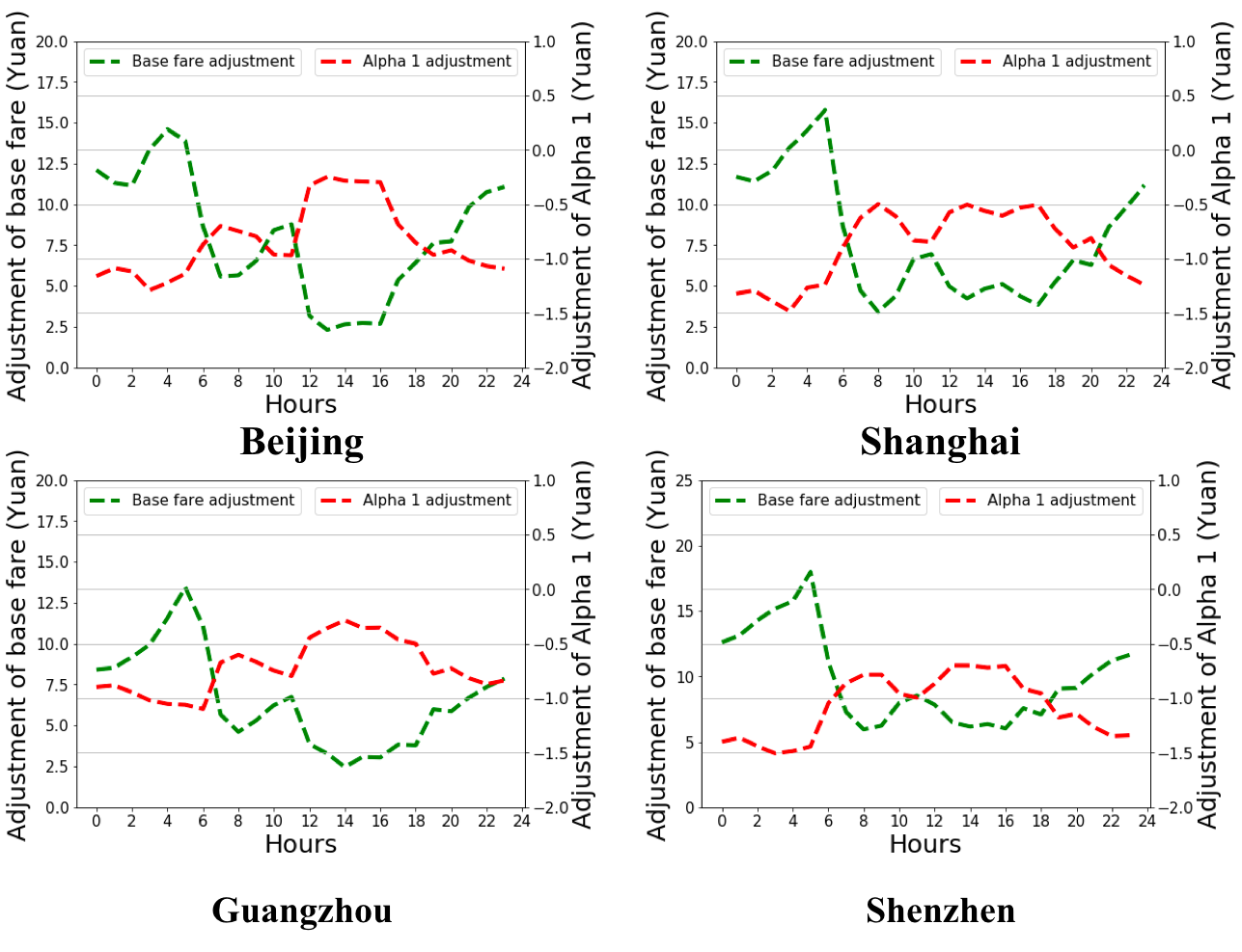}
\caption{$\Delta f_0$ and $\Delta \alpha_1$}
\label{Fig:approximation}
\end{figure}

$\Delta \alpha_1$ and $\Delta f_0$ can be calculated by solving equation \ref{Eqa:Pricing_equilibrium} and equation \ref{Eqa:Utility_equilibrium}, we further depict the results in Figure \ref{Fig:approximation}. Intuitively, it can be observed that all the $\Delta f_0$s will be positive while $\Delta \alpha_1$s will keeps negative in each time interval. In other words, we should rise $f_0$ and abate $\alpha_1$ to motivate drivers to avoid detours. Another interesting observation is that there exists a high peak of $f_0$ and a low peak of $\alpha_1$ in early hours (among 2:00-5:00), which implies that the detour utilities of early hours in reality are so high that should be urgently reduced. 

\section{Conclusion}

In this paper, we have investigated the problem how to eliminate the detour behaviors in E-hailing platforms, which is motivated by the fact that anomalous trajectories and pricing data can reveal many hidden “facts” about the human behaviors. To solve the problem, we propose a novel framework for detecting and analyzing the detour behaviors both in off-line database and among on-line trips. Applying our framework to real-world taxi data-set, a remarkable performance (AUC surpasses 0.98) has been achieved in off-line classification. Meanwhile, we further extend the off-line methods to on-line detection, a warning mechanism is introduced to remind drivers and an excellent precision (AUC surpasses 0.90) also has arrived in this phases. After conducting extensive experiments to verify the relationships between pricing regulations and detour behaviors, some quantitative pricing suggestions, including rising base fare and reducing distance-based fare rate, are provided to eliminate detour behaviors from the long term.

In the future, more real-world applications, such as road network changes mining and recommendation service correction, will be developed to validate our method. We believe that these value-added applications could benefit from our proposed method and significantly improve the level of taxi service.

\section*{Acknowledgment}

This work was done during internship of the second author in Map Department, DiDi chuxing. The authors would like to appreciate Siyuan Feng, a beautiful girl and an intelligent product manager in DiDi chuxing, and Yong Liu, a professional algorithm engineer in DiDi chuxing, for their kindness help and comprehensive advice.

\ifCLASSOPTIONcaptionsoff
  \newpage
\fi



%
\bibliographystyle{IEEEtran}
\bibliography{references}{}


\end{document}